\documentclass[letterpaper,12pt]{article}
\usepackage{tabularx} % extra features for tabular environment
\usepackage{amsmath}  % improve math presentation
\usepackage{graphicx} % takes care of graphic including machinery
\usepackage[margin=1in,letterpaper]{geometry} % decreases margins
\usepackage{setspace}
\usepackage{geometry}  % geometry package: left right top height width
\usepackage{caption}
\usepackage{cite} % takes care of citations
\usepackage[final]{hyperref} % adds hyper links inside the generated pdf file
\usepackage{tgbonum}
\usepackage{titlesec}
\usepackage{color,listings}
\hypersetup{
	colorlinks=true,       % false: boxed links; true: colored links
	linkcolor=blue,        % color of internal links
	citecolor=blue,        % color of links to bibliography
	filecolor=magenta,     % color of file links
	urlcolor=blue         
}
\definecolor{codegreen}{rgb}{0,0.6,0}
\definecolor{codegray}{rgb}{0.5,0.5,0.5}
\definecolor{codepurple}{rgb}{0.58,0,0.82}
\definecolor{backcolour}{rgb}{0.95,0.95,0.92}
\lstdefinestyle{mystyle}{
    backgroundcolor=\color{backcolour},   
    commentstyle=\color{codegreen},
    keywordstyle=\color{magenta},
    numberstyle=\tiny\color{codegray},
    stringstyle=\color{codepurple},
    basicstyle=\footnotesize,
    breakatwhitespace=false,         
    breaklines=true,                 
    captionpos=b,                    
    keepspaces=true,                 
    numbers=left,                    
    numbersep=5pt,                  
    showspaces=false,                
    showstringspaces=false,
    showtabs=false,                  
    tabsize=2
}
\begin{document}
\pagenumbering{roman}

\setcounter{page}{1}
\centerline{\bf{ Report on Data Pipeline Development}}
\vspace*{1\baselineskip}

\centerline{\bf{ for Grain Boundary Structures Classification }}
\vspace*{1\baselineskip}

\centerline{\bf{ By }}
\vspace*{1\baselineskip}

\centerline{\bf{ BINGXI LI }}
\centerline{\bf{ B.E. (Northwestern Polytechnical University) 2014 }}
\centerline{\bf{ M.S. (University of California, Davis) 2016 }}

\vspace*{1\baselineskip}
\centerline{\bf{ REPORT }}
\vspace*{1\baselineskip}

\centerline{\bf{ Submitted in partial satisfaction of the requirements for the degree of }}
\vspace*{1\baselineskip}

\centerline{\bf{ MASTER OF SCIENCE }}
\vspace*{1\baselineskip}

\centerline{\bf{ in }}
\vspace*{1\baselineskip}

\centerline{\bf{ Computer Science }}
\vspace*{1\baselineskip}

\centerline{\bf{ in the }}
\vspace*{1\baselineskip}

\centerline{\bf{ OFFICE OF GRADUATE STUDIES }}
\vspace*{1\baselineskip}

\centerline{\bf{ of the }}
\vspace*{1\baselineskip}

\centerline{\bf{ UNIVERSITY OF CALIFORNIA }}
\vspace*{1\baselineskip}

\centerline{\bf{ DAVIS }}
\vspace*{1\baselineskip}

\centerline{\bf{ Approved: }}
\vspace*{1\baselineskip}

\centerline{\bf{ }}
\centerline{\rule[--3mm]{10cm}{0.05em}}
\vspace*{-0.5\baselineskip}
\centerline{\bf{ Prof. Norman S. Matloff, Chair }}
\vspace*{1\baselineskip}

\centerline{\bf{ }}
\centerline{\rule[--3mm]{10cm}{0.05em}}
\vspace*{-0.5\baselineskip}
\centerline{\bf{ Prof. Francois Gygi }}
\vspace*{1\baselineskip}

\centerline{\bf{ Committee in Charge }}
\vspace*{1\baselineskip}
\centerline{\bf{ 2017 }}

\newpage
\onehalfspacing
% \fontfamily{qcr}
\vspace*{2\baselineskip}
\centerline{\rule[0pt]{6.5in}{0.1em}}
\vspace*{1\baselineskip}
\rightline{Bingxi Li}
\rightline{September 2017}
\rightline{Computer Science}
\vspace*{1\baselineskip}

\centerline{ Report on Data Pipeline Development }
\centerline{ for Grain Boundary Structures Classification }
\vspace*{1\baselineskip}
\centerline{\bf{\underline{ Abstract }}}
\vspace*{1\baselineskip}
\noindent \\
\\
Grain Boundaries govern many properties of polycrystalline materials, including the vast majority of engineering materials. Evolutionary algorithm can be applied to predict the grain boundary structures in different systems. However, the recognition and classification of thousands of predicted structures is a very challenging work for eye detection in terms of efficiency and accuracy. A data pipeline is developed to accelerate the classification and recognition of grain boundary structures predicted by Evolutionary Algorithm. The data pipeline has three main components including feature engineering of grain boundary structures, density-based clustering analysis and parallel K-Means clustering analysis. With this data pipeline, we could automate the structure analysis and develop better structural and physical understanding of grain boundaries.\\
\\
%\centerline{\rule[0pt]{6.5in}{0.1em}}
\vspace*{1\baselineskip}
\rightline{\it{September 29, 2017}}

\newpage
\listoffigures

\newpage
\tableofcontents

\newpage
\pagenumbering{arabic}
\section{Introduction}
\graphicspath{{./section1/Fig/}}
\subsection{Grain Boundary Structures}
Cleaner energy conversion and more efficient energy utilization generated increasing demand in the development of advanced metallic alloys and ceramics that can safely perform under extreme conditions. The performance and properties of these structural and functional materials are greatly determined by the existence of internal interfaces called grain boundaries (GB) during materials synthesis and processing. Therefore it is significant to understand the grain boundary structures and their influence on materials properties.\\
\\
\noindent In previous studies, interfacial structures existing in different states are called {\it complexions}\cite{dillon2007complexion,CANTWELL20141}. Complexion types are characterized by different amounts of impurity segregation. Different types of complexions including monolayer, bilayer, trilayer and thicker inter-granular films have been suggested\cite{dillon2007complexion}. Grain boundary complexions was firstly predicted and well studied by earlier theoretical researches with phase model for their roles in first-order and high order transition \cite{RICKMAN201388,tang2006grain,tang2006diffuse}. Experimental studies suggested a potential role of complexions transitions on abnormal grain growth in ceramics\cite{dillon2007complexion}, activated sintering\cite{luo1999origin}, and liquid metal embrittlement\cite{luo2011role}. Recently, GB complexions has been also applied to lattice dislocations, revealing the existence of new states of GB called linear complexions. All of the above studies suggested the importance of GB in determining the behaviors and structural properties of materials\cite{kaplan2015mechanism,kuzmina2015linear}.\\
\\
While the experimental investigation of the influence of grain boundary on materials properties is currently a highly active field\cite{baram2011nanometer,rheinheimer2015non,dillon2016importance,rohrer2016role}, the atomic details of these grain boundary phases remain unclear. It is extremely difficult to have direct experimental observations of interface phase transitions at high temperature by HRTEM due to inherent limitations\cite{merkle1987atomic}. Although many HRTEM studies of grain boundaries in doped metallic and ceramic materials showed grain boundary structures are similar inter-granular films of different thickness\cite{CANTWELL20141,dillon2007complexion,baram2011nanometer,park2003singular},these HRTEM images are of low resolution and cannot provide more detailed atomistic structure of these boundaries, which make the results and conclusions less convincing.\\
\\
Atomistic simulations is promising to predict atomic structure of interfaces and study their thermodynamic and kinetic properties. $\gamma$ surface method has been developed and well accepted as a common approach to build grain boundaries in atomistic simulations. It has been employed to study interfaces in a variety of materials for more than four decades. However an increasing number of recent studies suggested an alternative approaches of grain boundary construction and pointed out that $\gamma$-surface method is limited to predict true ground states.\\
\\
For example, a recent investigation of two high angle boundaries $\Sigma$5(210)[001] and $\Sigma$5(310)[001] in Cu demonstrated the root cause of missing transformations is the inadequate simulation methodology with constant number of atoms and periodic boundary conditions. An alternative high-temperature anneals of these boundaries connected to open surfaces allowed the variation of atom numbers in grain boundary to achieve lower free energy states\cite{frolov2015segregation,frolov2016phase}. The simulations revealed multiple new grain boundary phases of the boundaries are impacted by different atomic densities and demonstrated fully reversible first-order transitions induced by temperature, changes in chemical compositions and point defects. This ingenious modeling approach demonstrated phase behavior of two special high-angle boundaries that have been extensively investigated in the past still missed entire phenomenon which are overlooked by modeling with previous restrictive simulation methodology.\\
\\
Recognizing the limitations of current modeling capabilities and the obstacle to observing grain boundary phase transitions, a robust computational tool is required to predict complex grain boundary structures.

\subsection{Evolutionary Search}
\begin{figure}[ht!] 
\centering
\captionsetup{justification=centering}
\includegraphics[width=0.8\textwidth]{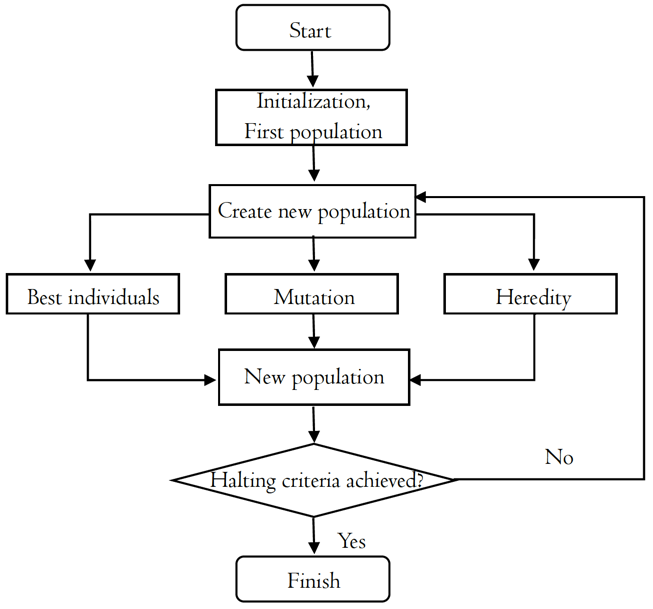}
\caption{\label{fig:generalUSPEX}A scheme of USPEX evolutionary prediction.}
\end{figure}

In recent years, there have been significant advances in predicting the structures from first-principles calculations. Among them, evolutionary algorithm has proved to be extremely powerful in different systems including bulk crystals\cite{oganov2006crystal}, 2D crystals\cite{zhou2014semimetallic}, surfaces\cite{zhu2013evolutionary}, polymers\cite{zhu2014predicting} and clusters\cite{lyakhov2013new}, etc. It is very promising to extend the method to predict grain boundary structures.\\
\\
A few pioneering works have been reported in the literature\cite{von2006structures,chua2010genetic,zhang2009finding}. However there are some limiting facts about these previous work. For instance, Chua et al developed a genetic algorithm to study the non-stoichiometric grain boundaries of $SrTiO_3 $\cite{chua2010genetic}. It can only be applied for a system with fixed number of atoms and super--cell size. In my research, a evolutionary algorithm based software \textbf{U}niversal \textbf{S}tructure \textbf{P}redictor: \textbf{E}volutionary \textbf{X}tallography (\textbf{USPEX}) \cite{oganov2006crystal} is used. It enables the automated exploration of GB structures with variable number of atoms and cell sizes in higher dimensional space. Here is an outline steps and a scheme Fig~\ref{fig:generalUSPEX} of USPEX prediction.\\
\\
The 7 steps are:
\begin{enumerate}
\item Prepare adequate representations for the problem: a one-to-one correspondence between the point in the search space and a set of numbers.
\item Initiate the first generation, which consists of a set of points in the search space satisfying the constraints of the problem.
\item Determine the quality of each member of the population with the fitness function.
\item Select the ``best'' member of the population as the parents and apply the variation to create the new points (offspring).
\item Evaluate the new member of the new population.
\item Select the ``best'' member of offspring to form the new generation of the population.
\item Repeat step 4 to 6 until reaching the halting criteria.
\end{enumerate}

\noindent It is well-known that the complexity exponentially increases with the growing dimensionality. Therefore to ensure efficient sampling, it is significant to balance between individual quality and population diversity. Naive random structure initialization or variation operation will only lead to disordered-like structures but with close energetics. To address this challenge, the implementation in USPEX utilizes coarse-grained modeling and defines the simplified representations when generating structures. Some key representations including symmetry, vibrational modes and degree of local order are used here.\\
\\
With the powerful prediction tool based on evolutionary algorithms, we are able to predicts structures of interfaces. In the iterations of Fig~\ref{fig:generalUSPEX}, USPEX generates a population of grain boundary structures and optimize them over several generations to predict configurations of low-energetics. During the evolutionary convergence process, complex and diverse structures with different atomic densities are sampled by operations of heredity and mutation which involve atomic rearrangements like addition and removal of atoms from the grain boundary core.

\subsection{Motivations of Data Pipeline Development}
In simulation, the evolutionary algorithm generates thousands of GB structures in every prediction run to Cu system with different tilting angle $\theta$, ranging from 11.42$^\circ$ to 79.61$^\circ$. Here are challenges in analysis over such a large number of GB structures.\\
\\
\noindent {\bf Undefined Problems}\qquad Most of the GB structures are first time to observe and very few studies can be referenced when classifying and understanding these GB structures. So there is no clear and complete definition of GB structures in Cu system with different tilting angles. Besides, most of these structures are disordered or mixed with different complexions. Some structures that exist in Cu system with certain angles miss in other system with different tilting angles. This makes the existence of GB structures inconsistent through a wide range of tilting angles. A convincing way to represent and describe GB structures is therefore necessary.\\
\\
\noindent{\bf Lacking of Robust Classification}\qquad It is quite time-consuming to classify over thousands of structures through eye detection. This process needs laboring work and detailed check and comparisons for multiple times, which greatly reduce the efficiency of GB discovery.\\
\\
To solve the above challenges, solutions are proposed as follows:\\
\\
{\bf Mapping GB Structure in Lower Dimensional Space}\qquad It is clear that the GB structure is originally represented in 3$N$-dimensional space, where $N$ is the number of atoms in GB structures. Therefore a technique to reduce dimension is required. In my work, every GB structure is represented in a feature space composed by structural and mechanical properties including excessive free energy, excessive volume, stress tensors, Steihardt order parameters (Q$_4$, Q$_6$, Q$_8$, Q$_{12}$). Through this feature engineering work, a problem in 3$N$-dimensional space is then reduced into one in 8-dimensional feature space or properties space.\\
\\
{\bf Density Based Clustering Algorithm}\qquad The categorization of the GB structures is actually the learning of non-labeled data, which is equally a unsupervised learning problem. Clustering algorithm is very promising for this type of problem. However, conventional clustering algorithm like K-means clustering algorithm requires prior knowledge to select the value for K, which limits the power of this algorithm. A density based clustering algorithm proposed by Rodriguez, Alex and Laio, Alessandro can yield reasonable K value without any domain knowledge. All the data can be clustered with $O(1)$ iterations, namely without self-consistent process.\\
\\
To integrate the feature engineering and density based clustering algorithm, I developed the data pipeline to automate the whole analysis. It has been proved to be very helpful and useful in the GB structures analysis.

\section{Components of Data Pipeline}
In this section, I would like to introduce the ideas behind every component of the data pipeline.
\graphicspath{{./section2/Fig/}}
\subsection{Feature Engineering of Grain Boundary Structures}
Firstly, I will give a brief review of the grain boundary structure model and the evolutionary prediction.
\subsubsection{Evolutionary Predicted Structures}
\begin{figure}[ht!] 
\centering
\captionsetup{justification=centering}
\includegraphics[width=0.8\textwidth]{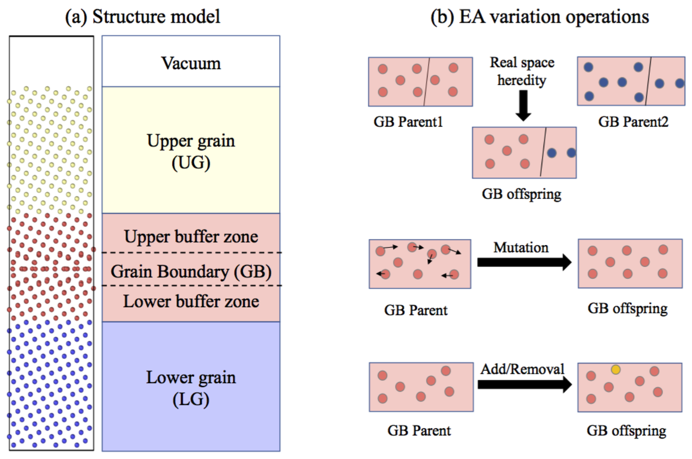}
\caption{\label{fig:gbmodel}Structure model and evolutionary operations for prediction.}
\end{figure}

\noindent In Fig~\ref{fig:gbmodel} The GB model used for into three different regions, the region of upper grain (UG) and lower grain (LG), and grain boundary (GB). The evolutionary algorithm adopts concepts from evolutionary biology based on populations, selection, reproduction by heredity and mutation to optimize the individual with highest fitness. The code generates a population of grain boundary structures and improves them over several generations to predict low-energy configurations. During the evolution complex and diverse structures with different atomic densities are sampled by operations of heredity and mutation which involve atomic rearrangements as well as addition and removal of atoms from the grain boundary core.\\

\noindent The evolutionary algorithm adopts three operations populations, selection, reproduction by heredity and mutation to optimize the individual with highest fitness. The code generates a population of grain boundary structures and improves them over several generations to predict low-energy configurations. During the evolution complex and diverse structures with different atomic densities are sampled by operations of heredity and mutation which involve atomic rearrangements like addition and removal of atoms from the grain boundary core.\\
\\
\noindent The Evolutionary Prediction finally output different configurations of grain boundary structures. And these structures are represented as set of atomic coordinates in 3-dimension Cartesian space. Thus, every structure is a data point in 3N-dimensional space, where N is the number of atoms that configure the structure.

\subsubsection{Construction of Feature Space}
\noindent Eight excessive properties are calculated from the GB region and bulk region shown in Fig~\ref{fig:gbmodel}. The excessive properties are excessive free energy $[\gamma]_N$, atomic volume $[V]_N$, stress $[\tau]_{11}$, stress $[\tau]_{22}$ and Steinhardt order parameter $[Q_4]_N$, $[Q_6]_N$, $[Q_8]_N$, $[Q_{12}]_N$\cite{steinhardt1983bond}.\\
\\
\noindent In a single component system, grain boundary free energy $\gamma_{GB}$ is given with following equation.
\begin{equation}
\begin{split}
\gamma A & = E - TS - \sigma_{33}V = [E]_N - T[S]_N - \sigma_{33}[V]_N
\end{split}
\end{equation}
\noindent where $[Z]_X$ are grain boundary excess properties expressed with Cahn’s determinants. According to the adsorptions equation\cite{frolov2012thermodynamics}, grain boundary free energy is also a function of temperature, stress and lateral strain.
\begin{equation}
\begin{split}
d(\gamma A) = -[S]_N\ dT-[V]_N\ d\sigma_{33} + \tau_{ij} A\ de_{ij}
\end{split}
\end{equation}
\noindent where i,j = 1,2. At 0K, the excess volume $[V]_N$ and two components of grain boundary stress $\tau_{11}$ and
$\tau_{22}$ per unit of grain boundary area can be computed as,
\begin{equation}
\begin{split}
[V]_N &= \frac{1}{A}(V-V^{bulk}\frac{N}{N^{bulk}})\\
\tau_{11} &=[\sigma_{11}V]_n = \frac{1}{A}(\sigma_{11}V - \sigma_{11}^{bulk}V^{bulk}\frac{N}{N_{bulk}})\\
\tau_{22} &=[\sigma_{22}V]_n = \frac{1}{A}(\sigma_{22}V - \sigma_{22}^{bulk}V^{bulk}\frac{N}{N_{bulk}})
\end{split}
\end{equation}
\noindent Notice that $\frac{V^{bulk}}{N_{bulk}} = \Omega$ is a volume per atom in the bulk. In atomistic simulations volume occupied by each atom is calculated by LAMMPS with the Voronoi construction\cite{plimpton1995fast}. The product $\sigma_{ij}V$ for each atom can also be calculated by LAMMS. In our calculations bulk stresses are zero within the numerical accuracy.\\
\\
Besides the four features described above, excess amounts of Steinhardt order parameters of grain boundary is also used, including $Q_4$, $Q_6$, $Q_8$ and $Q_{12}$\cite{steinhardt1983bond}. These parameters per atom are calculated within LAMMPS\cite{plimpton1995fast}. The equation for this calculation is,
\begin{equation}
\begin{split}
[Q]N = \frac{1}{A}(Q-Q^{bulk}\frac{N}{N_{bulk}})
\end{split}
\end{equation}
\noindent where $Q = \sum_{i=1}^{N}Q^i$ is the total amount of the order parameter per atom in a region enclosing the grain bulk and containing N atoms, $\frac{Q^{bulk}}{N^{bulk}}$ is the value of this order parameter per atom in the bulk, where Q is one of the Q4, Q6, Q8 or Q12.\\
\\
\noindent The above excessive properties calculation takes the grain boundary structure in form of atomic coordinates as input. Therefore after the calculations to 8 excess properties for every structure, we finally map each structure in 3N-dimensional space into a data point in 8-dimensional feature space.

\subsection{Clustering Analysis}
\subsubsection{Density Based Clustering Analysis}
With the above mapping techniques, GB structure can be represented as a vector f in the feature space, where 
\begin{equation}
\begin{split}
f=([\gamma]_N, [V]_N), \tau^{11}_N, \tau^{22}_N, [Q_4]_N, [Q_6]_N, [Q_8]_N, [Q_{12}]_N 
\end{split}
\end{equation}

\noindent The density based clustering algorithm\cite{rodriguez2014clustering} is actually a visualized methodology. It firstly remap the vector for data i in the feature space onto a two dimensional plot where the x, y axis are density $\rho_i$ and $\delta_i$, which measures the minimum distance between the point i and any other point with higher density.\\
\\
\noindent Therefore for each data point i, we need to compute two quantities: its local density $\rho_i$ and its distance $\delta_i$ from points of higher density. Both these quantities depend on the distances $d_{ij}$ between data points, which are assumed to satisfy the triangular inequality. For the local density $\rho_i$ of data point i, it is defined as the Euclidean distance between structure i, j in this abstract space.
\begin{equation}
\begin{split}
d_{ij} = ||f^i-f^j||_2 = \sqrt{\sum_{n=1}^{8}(f_n^i-f_n^j)^2}
\end{split}
\end{equation}

\noindent The local density $\rho_i$ of data point i is defined as,
\begin{equation}
\begin{split}
\rho_i = \sum_j\chi(d_{ij}-d_c)
\end{split}
\end{equation}

\noindent where $\chi(x)=1$ if $x<0$ and $\chi(x)=0$ otherwise, $d_c$ is a cutoff distance. It can be simply expressed as the number of points that are closer than $d_c$ to point i. The algorithm is only robust with a good choice of $d_c$. And the choice of $d_c$ can be well derived with following visualization way.\\
\\
\noindent For the $\delta_i$, it is assigned with following equation,
\begin{equation}
\begin{split}
\delta_i = \min_{j:\rho_j>\rho_i}(d_{ij}).
\end{split}
\end{equation}

\noindent while for the point with highest density, we conventionally take $\delta_i = \max_j(d_{ij})$. Note that di is much larger than the typical nearest neighbor distance only for points that are local or global maxima in the density. Thus, cluster centers are recognized as points for which the value of $\delta_i$ is anomalously large.\\

\begin{figure}[ht!] 
\centering
%\captionsetup{justification=centering}
\includegraphics[width=1.0\textwidth]{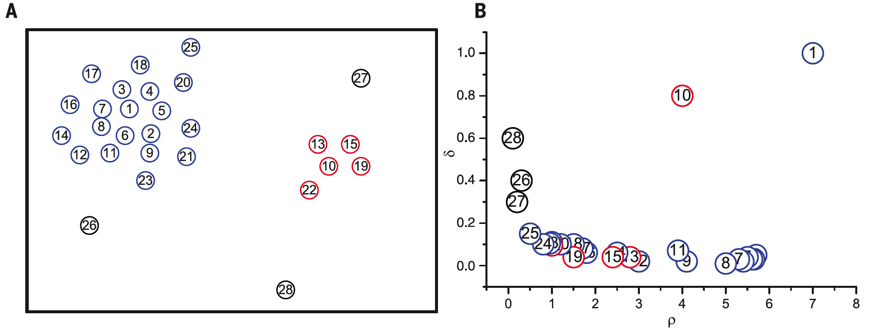}
\caption{\label{fig:dbc} Two dimensional visualization of the algorithm. (A) Point distribution. Data points are ranked in order of decreasing density. (B) Decision graph for the data in (A). Different colors correspond to different clusters.\cite{rodriguez2014clustering}}
\end{figure}

\noindent In Fig~\ref{fig:dbc}, the algorithm in two dimension is described with an example of 28 data points. Fig~\ref{fig:dbc}A contains 28 points. It can be seen that points 1 and 10 are the density maxima, which we identify as cluster centers. Fig~\ref{fig:dbc} shows the points distribution on the plot of $\rho$ and $\delta$. This representation is a decision graph. From the graph, we could see that the value $\rho$ of points 9 and 10 is similar while their values of $\delta$ is very different. Point 9 belongs to the cluster of centering around point 1, and several other points with a higher $\rho$ are very close to it, whereas the nearest neighbor of higher density of point 10 belongs to another cluster. Hence, as anticipated, the only points of high $\rho$ and relatively high $\delta$ are the cluster centers. Points 26, 27, and 28 have a relatively high d but with a low r. This is because they are isolated. They can be regarded as clusters of a single point or outliers\cite{rodriguez2014clustering}.\\
\\
\noindent A great advantage of this algorithm is that after assigning the cluster centers, each remaining point is assigned to the same cluster as its nearest neighbor of higher density. Therefore, the decision graph can be computed within $O(n^2)$ time complexity and $O(n)$ space complexity, where n is the number of GB structures. The cluster assignment is performed in a constant step, in contrast with other clustering algorithms which usually need to optimize the objective function iteratively\cite{le1972proceedings,mclachlan1997wiley}.

\subsubsection{Parallel K-means Clustering Functionality}
K-means clustering is a popular method for cluster analysis in data mining. It aims to partition n observations into k clusters in which each observation belongs to the cluster with the nearest mean, serving as a prototype of the cluster. As a result of K-means clustering, the data space will be partitioned into Voronoi cells.\\
\\
The most common algorithm uses an iterative refinement technique. The popular one is often referred as Lloyd's algorithm. It starts with an initial set of k means $m_1^{(1)}, \ldots, m_k^{(1)}$. Then the algorithm proceeds by alternating between two following steps \cite{mclachlan1997wiley}:
\begin{enumerate}
\item {\bf Assignment step:} Assign each observation to the cluster whose mean has the least squared Euclidean distance. This mathematically means partitioning the observations according to the Voronoi diagram generated by the means.
\begin{equation}
\begin{split}
S_i^{(t)} = \{x_p : ||x_p-m_i^{(t)}||_2^2\leq||x_p-m_j^{(t)}||_2^2\ \forall j, 1\leq j \leq k\}
\end{split}
\end{equation}
\item {\bf Update step:} Calculate the new means to be the centroids of the observations in the new clusters.
\begin{equation}
\begin{split}
m_i^{(t+1)} = \frac{1}{|S_i^{(t)}|}\sum_{x_j\in S_i^{(t)}}x_j
\end{split}
\end{equation}
\end{enumerate}

\noindent The algorithm has converged when the assignments no longer change. However it does not guarantee that the optimum found using this algorithm.\\
\\
\noindent{\bf Motivations to Explore Parallelism in K-Means Algorithm}\\
\noindent K-means algorithm is a very simple and powerful method for clustering analysis. However, finding the optimal solution to the k-means clustering problem for observations in d dimensions is a NP-hard problem\cite{inaba1994applications}. There were a variety of heuristic algorithms such as Lloyd's algorithm given above are generally used. The running complexity of Lloyd's algorithm is ${O(nkdi)}$\cite{christopher2008introduction}, where n is the number of entities represented as d-dimensional vectors to cluster and k the number of clusters, i being the number of iterations needed until convergence.\\
\\
The number of iterations until convergence is only small, on data that have a cluster structure. Results can improve slightly after the first dozen iterations. Lloyd's algorithm is therefore often considered to be of "linear" complexity in practice. But in worst case, it is superpolynomial\cite{arthur2006slow}. Another thing that makes the case even worse is the large quantity of data. We are in a world of big data today. Most of clustering algorithms unavoidably face the challenges raised by massive data. Through the above analysis, we see the rapid complexity growth of this algorithm when clustering over data at large scale. Therefore it will be very promising to explore the potential parallelism in current algorithm and accelerate it on powerful parallel platform.\\
\\
\noindent{\bf Parallelism in K-Means Algorithm}\\
The parallelism is obvious in the assignment step. The serial version will take $O(N^2)$ time complexity. However the distance computation have data independence to use, which means the assignment of data $i$ can be done simultaneously with data $j$. To explore the data-level parallelism, some parallel computation model like CUDA C and OpenACC are very favorable. Compared to CUDA, OpenACC requires much less programming effort to parallelize serial code.

\section{Implementation}
This section discusses some details of the data pipeline implementation.
\graphicspath{{./section3/Fig/}}
\subsection{GB structure calculations at 0 K} 
Each grain boundary was generated from 3-5 independent evolutionary searches. Each search evolves up to fifty generations. The search explores different atomic densities ranging from 0 to 1 measured as a fraction of number of atoms found in one bulk atomic plane parallel to the grain boundary. A typical run explores the structures ranging from 500 to 5000 atoms for the entire model and 30 to 300 atoms for the GB region. The different grain boundary areas of each grain boundary are explored by replicating the smallest possible cross-section up to 25 times. LAMMPS code\cite{plimpton1995fast} was used to evaluate the energy of every generated configurations.\\
\\
The properties calculation are performed in batch over thousands of structures. Task monitor is run with Linux Crontab service for periodically checking of job status and submission. This helps accelerate the computing process with some parallelism.

\subsection{Density Clustering}
An R CRAN package implementing the density based clustering algorithm by Alex Rodriguez and Alessandro Laio\cite{rodriguez2014clustering} is incorporated in this data pipeline. This package `densityClust' provides robust tools to generate the initial $\rho$ and $\delta$ values for each observation as well as to use these thresholds to assign observations to clusters. Since this is done in two passes, it is free to reassign observations to clusters using a new set of $\rho$ and $\delta$ thresholds, without needing to recalculate everything.

\subsection{Parallel Implementation of K-Means Algorithm}
\begin{figure}[ht!] 
\centering
\captionsetup{justification=centering}
\includegraphics[width=0.8\textwidth]{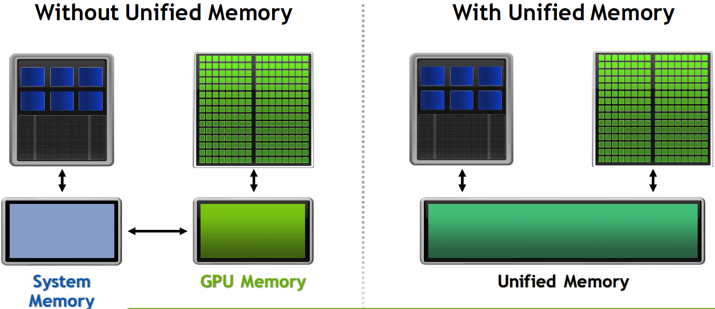}
\caption{\label{fig:programmingmodel}The heterogeneous architecture for OpenACC (\href{https://www.OpenACC.org/}{https://www.OpenACC.org/}).}
\end{figure}

The parallel K-Means algorithm is implemented with OpenACC. OpenACC is a user-driven directive-based performance-portable parallel programming model designed to help scientists and engineers to porting their codes over a wide-variety of heterogeneous platform with significantly less programming effort than that with a low-level model \href{https://www.OpenACC.org/}{https://www.OpenACC.org/}).\\
\\
OpenACC takes use of the computing power of GPU and it is CUDA related. This programming model is therefore based on the heterogeneous architecture. The hardware model is shown in Fig.~\ref{fig:programmingmodel}.

\noindent Without unified memory, the data needs to be copied from CPU end to GPU end through PCIe. With this programming model, the K-means algorithm can be accelerated in following porting cycles.

\subsubsection{3 Steps to Accelerate with OpenACC}
Acceleration of a serial program can be divided into 3 steps \textbf{Analysis}, \textbf{Parallelization}, \textbf{Optimization}. The analysis focuses on the most time-consuming part of a serial program. Usually it is multiple loops over massive data. Parallelization adds on appropriate directives to highlight where the parallelism lies. Optimization usually focuses on the data movement between CPU and GPU end.

\subsubsection{Parallel Directives}
\noindent In previous section 2, we have already analyzed the parallelism in K-means Clustering algorithm. Therefore we should use the methods, the \textbf{kernels} and \textbf{parallel} directives, provided by OpenACC to parallelize the corresponding area.\\
\\
\textbf{[kernels]} \ \ Using kernels simply leave the parallelism initialization to the compiler. One can use the kernels to automatically parse the region, and analyze potential parallelization capabilities. The kernels construct is very easy to use. Here is a simple example of using kernel directive.
\begin{lstlisting}[firstnumber=9]
#pragma acc kernels
{
	for(i = 0; i < n; i++){
		y[i] = 0;
		x[i] = i+1;
	}
	for(i = 0; i < n; i++){
		y[i] = 2 * x[i];
	}
}
\end{lstlisting}
By looking at the compile information, we can tell that the parallelism is automatically initialized by the complier. The line 9 shows the data transfer of matrix X and Y from host to device memory. The compiler automatically recognizes the parallelism in the ``for'' loop starting from line 11 and generates a parallel loop operation correspondingly. The gang in line 11 is a thread block and the thread in CUDA is called vector in OpenACC.\\
\\
\noindent However, kernels directives does not guarantee the initialization of parallel loop. The compiler is poor at determining data dependency. It will not parallelize the loop that fails the dependency analysis. Then it will be developers' responsibilities to parallel.
\begin{lstlisting}
9, Generating copyout(x[:],y[:])
11, Loop is parallelizable
Accelerator kernel generated
Generating Tesla code
11, #pragma acc loop gang, vector(32) /* blockIdx.x threadIdx.x */
15, Loop is parallelizable
Accelerator kernel generated
Generating Tesla code
15,#pragma acc loop gang, vector(32) /* blockIdx.x threadIdx.x */
\end{lstlisting}

\noindent\textbf{[parallel loop]} \ \ As mentioned above, using ``parallel loop'' transfers the responsibility to identify the parallelism to the developer. The ``parallel'' notifies the complier to generates one or more parallel gangs, which execute redundantly and the ``loop'' directive informs the compiler which loop to parallelize. Sometimes the two directives can be used separately. Namely, the ``loop'' can be used singly before the loops which you believe have parallelism inside ``parallel'' region.\\
\\
\noindent A big advantage to parallelize the loops of ``parallel loop'' over ``kernels'' is shown as follows. When the ``kernels'' fails to recognize parallel loop for potential pointer aliasing issue, the developer could still use the ``parallel'' and ``loop'' to parallelize it. When using ``kernels'', we find the failure to recognize parallel loop according to the profiling information. 
\begin{lstlisting}[firstnumber=4]
void setArrays(int *x, int *y, int n){
    int i;
    #pragma acc kernels
    {
        for(i = 0; i < n; i++){
            y[i] = 0;
            x[i] = i+1;
        }
        for(i = 0; i < n; i++){
            y[i] = 2 * x[i];
        }
    }
}
\end{lstlisting}
\noindent The line 8 and 12 loops are not recognized as parallel loop by the compiler for potential pointer aliasing issue. The pointer aliasing refers to two arrays shares the same memory. The compiler can not determine it at compile time. Therefore the loops are kept for serial implementation for being safe.
\begin{lstlisting}
6, Generating copyout(x[:n],y[:n])
8, Complex loop carried dependence of y-> prevents parallelization
Loop carried dependence of x-> prevents parallelization
Loop carried backward dependence of x-> prevents vectorization
Accelerator scalar kernel generated
Accelerator kernel generated
12, Complex loop carried dependence of x-> prevents parallelization
Loop carried dependence of y-> prevents parallelization
Loop carried backward dependence of y-> prevents vectorization
Accelerator scalar kernel generated
Accelerator kernel generated
\end{lstlisting}

\noindent If developer is sure about no pointer aliasing issue, developer can use ``parallel loop'' to enforce the parallelism to the compiler at this loop as follows.
\begin{lstlisting}[firstnumber=4]
void setArrays(int *x, int *y, int n){
    int i;

    #pragma acc parallel loop
    for(i = 0; i < n; i++){
        y[i] = 0;
        x[i] = i+1;
    }
    #pragma acc parallel loop
    for(i = 0; i < n; i++){
        y[i] = 2 * x[i];
    }
}
\end{lstlisting}
\noindent The loop at line 8 and 13 are therefore implemented parallel.
\begin{lstlisting}
7, Accelerator kernel generated
Generating Tesla code
8, #pragma acc loop gang, vector(128) /* blockIdx.x threadIdx.x */
7, Generating copyout(x[:],y[:])
12, Accelerator kernel generated
Generating Tesla code
13, #pragma acc loop gang, vector(128) /* blockIdx.x threadIdx.x */
12, Generating copyout(y[:])
Generating copyin(x[:])
\end{lstlisting}

\noindent Besides, the pointer aliasing issue can be avoided with \textbf{restrict} keyword. With restrict before array y in declaration, ``int *restrict y'', the loops can also be recognized by compiler using ``kernels''.

\subsubsection{Optimization Techniques}
\noindent Optimization of OpenACC techniques is a broad topic. Here we mainly discussed the data directives in OpenACC and how it could be used in K-means algorithm. Commonly used data directives are given as follows.
\begin{enumerate}
    \item \textbf{copyin}: allocate memory on GPU and copies data from host to GPU when entering region.
    \item \textbf{copyout}: allocate memory on GPU and copies data to the host when existing region.
    \item \textbf{copy}: allocate memory on GPU and copies data from host to GPU when entering region and copies data to the host when exiting region.(Structured Only)
    \item \textbf{create}: allocate memory on GPU but not copy.
    \item \textbf{delete}: deallocate memory on the GPU without copying. (Unstructured Only)
    \item \textbf{present}: data is already present on GPU from another containing data region. 
\end{enumerate}

\noindent An obvious use of data directives in K-means algorithm is the \textbf{copyin} for source data. Since the data points will not be updated during the computation. Therefore, we could reduce the memory traffic without copying out the source data to the CPU end.\\
\\
\noindent For more materials about OpenACC programming, please refer to OpenACC official recommendations at \href{https://www.OpenACC.org/resources}{https://www.OpenACC.org/resources}.
\section{Results}
\graphicspath{{./section4/Fig/}}
\subsection{Grain boundary energy as a function of angle and atomic density}
\begin{figure}[htbp] 
\centering
\captionsetup{justification=centering}
\includegraphics[width=0.8\textwidth]{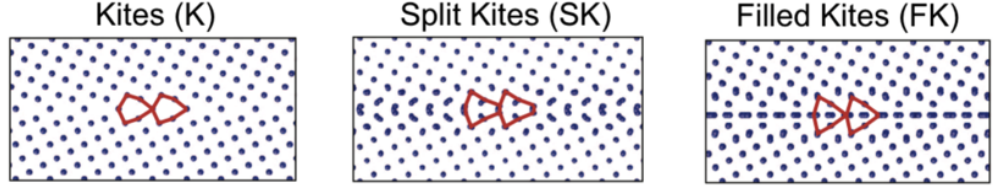}
\caption{\label{fig:kites}Representative structures from three popular kites family of GB structures\cite{submittedbl}.}
\end{figure}

Through the evolutionary search over a wide tile angle range($11.42^\circ\leq \theta\leq 79.61^\circ$), different GB structures are discovered. There are three popular types of GB structures in Fig.~\ref{fig:kites}. These representative structures are different from each other in the structure unit called kite.

\begin{figure}[htbp] 
\centering
%\captionsetup{justification=centering}
\includegraphics[width=0.8\textwidth]{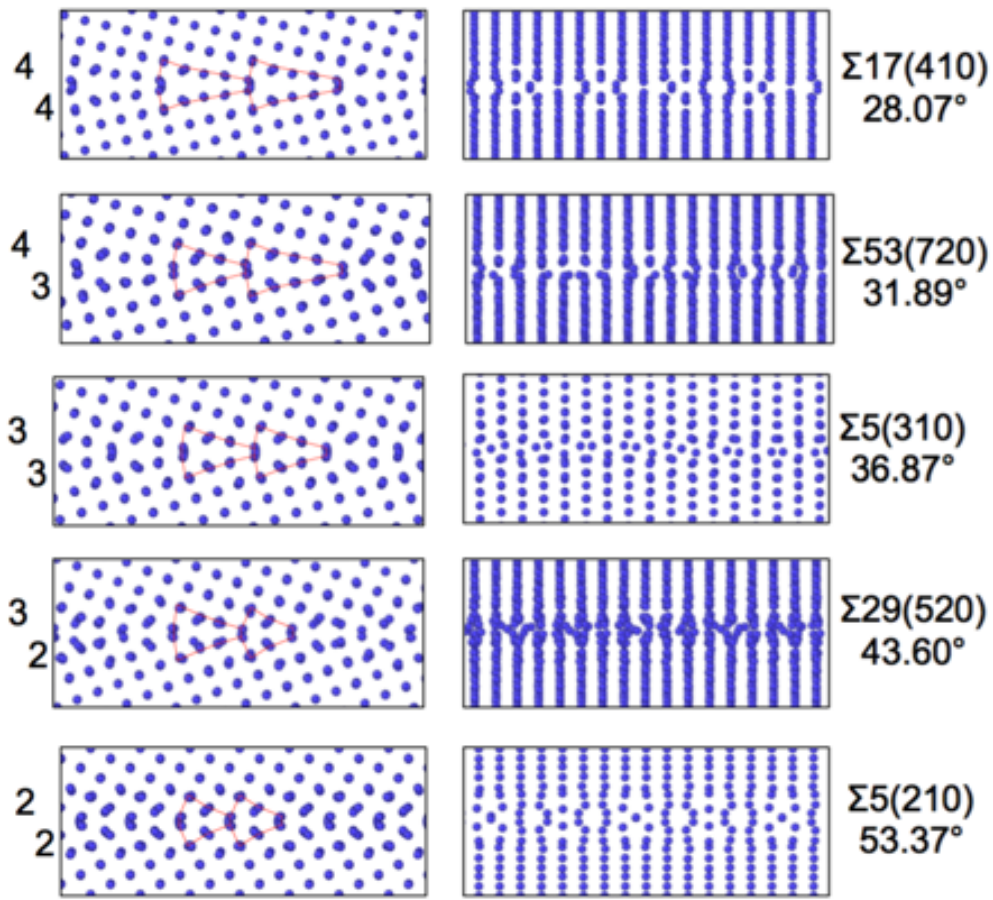}
\caption{\label{fig:sk}Split Kite Family. Split kites of five representative boundaries predicted by the evolutionary search and clustering analysis. These GB structures are viewed parallel to [001] tilt axis in left column and normal to it in right column.}
\end{figure}

\noindent Structures from three kites families also have their variations over a wide range of tilt angles. The mutations of split kites(SK) structure are shown in Fig.~\ref{fig:sk}. Split Kites have higher atomic density compared to Kites as extra atoms occupy interstitial positions between [001] planes.\\

\begin{figure}[htbp!] 
\centering
%\captionsetup{justification=centering}
\includegraphics[width=0.8\textwidth]{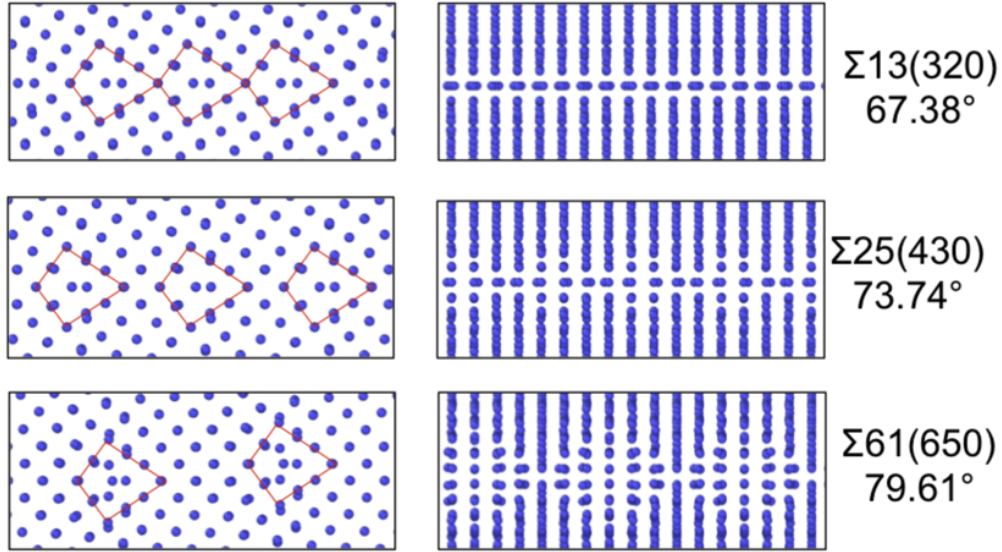}
\caption{\label{fig:ek}Extended Kite family. Three representative Extended Kite phases are predicted by the evolutionary search at 0 K. The misorientation angles are indicated on the figure. For each misorientation GB structures as viewed parallel to the [001] tilt axis in left and normal to it in right.}
\end{figure}

\noindent Another very important kite family in the high tilt angel range is Extended Kite. Extended Kites have higher atomic density compare to Kites, which correspond to half of the atomic plane in Fig~\ref{fig:ek}. The structural units are outlined and change their separation with the increasing misorientation angle.\\

\begin{figure}[ht!] 
\centering
\includegraphics[width=0.5\textwidth]{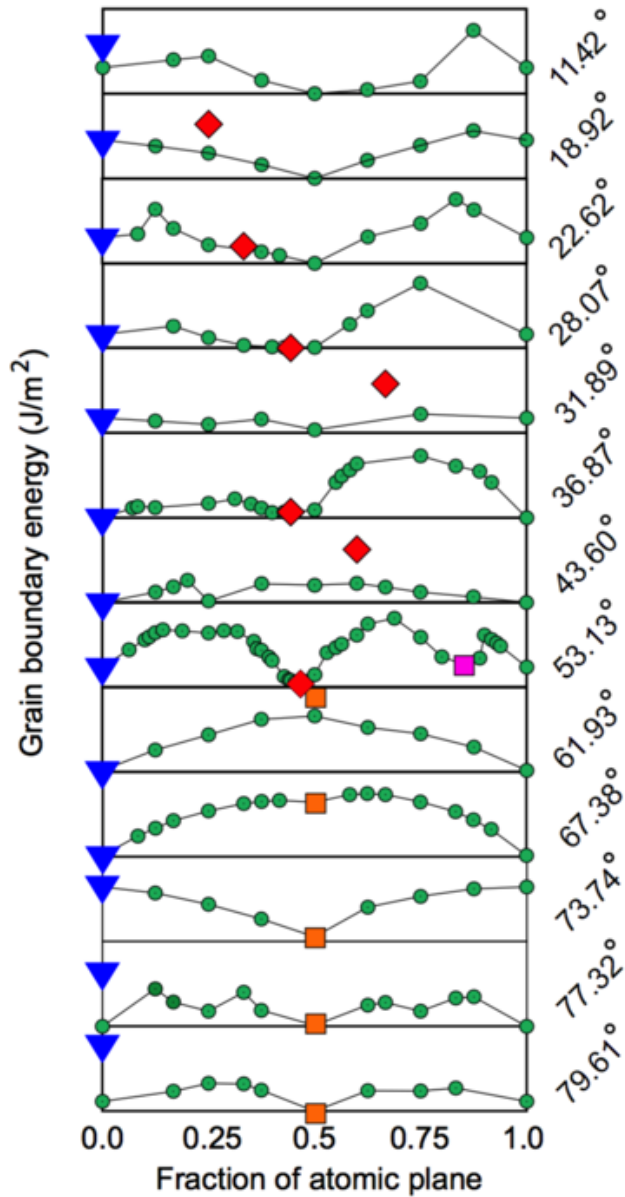}
\caption{\label{fig:energymap}Evolutionary search and clustering identify new ground states and multiple grain boundary phases. The search explores different atomic densities and finds low-energy grain boundary configurations (green circles) ignored by the conventional methodology. With same tilt angle $\theta$ atomic fractions and energies of each grain boundary at 0 K are colored by blue triangles for Kite family , red diamonds for Split Kite family and orange squares for Extended Kite family.}
\end{figure}

\noindent By varying misorientation angle and atomic density, the energy map of grain boundary phases can be derived easily. It is shown in Fig~\ref{fig:energymap} that the Split Kite dominates the lower-angle grain boundary while most of grain boundary structures with higher tilt angles are Extended Kite. In the middle between high and low angle region, the Kite structure has lowest energy. The structures on convex hulls of Fig~\ref{fig:energymap} are determined through eye detection for its small quantity.

\subsection{Clustering results for the grain boundary}
\begin{figure}[htbp] 
\centering
\includegraphics[width=0.7\textwidth]{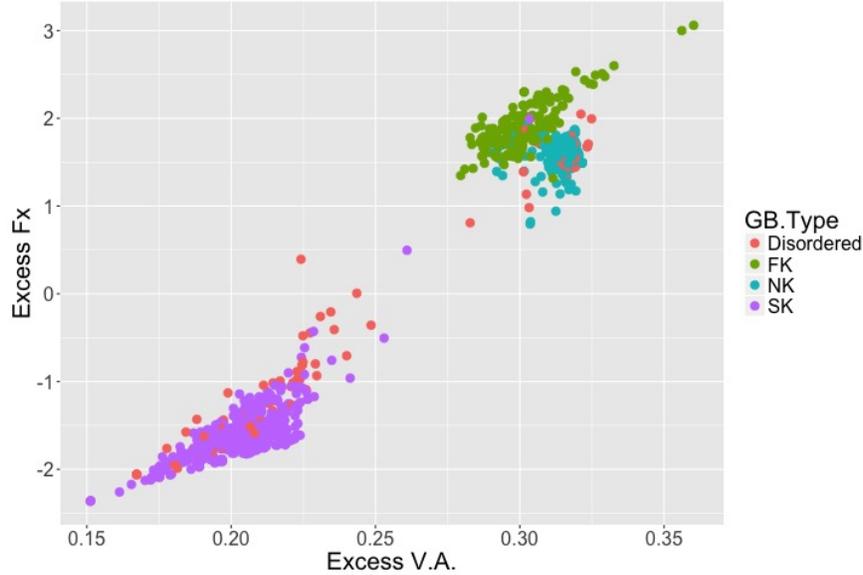}
\caption{\label{fig:210FxV}Eye detection of $\sum$ 5(210) in the 2-dimensional space of excess stress and volume, where the direction of excess stress $F_x$ is parallel to tilt axis. The orange dots correspond to distorted structures while the green, blue and purple are for Filled Kite(FK), Normal Kite(K or NK) and Split Kite(SK) respectively.}
\end{figure}

\noindent The evolutionary search is a very efficient methodology to discover grain boundary structures automatically. It usually will sample thousands of different grain boundary structures, which lead to the challenge in analyzing such a large amount of data. The Fig~\ref{fig:210FxV} shows the clustering results of around 1000 structures in the 2-dimensional space of excess stress and volume. The direction of the excess stress $F_x$ is parallel to the tilt axis. It is clear to see that the excess stress and volume are effective order parameter pair to distinguish different kite structures. However, this clustering behavior is discovered through laboring and time-consuming eye-detection process. It is therefore necessary to improve this process with an intelligent algorithm to find good order parameters and to detect the clusters.\\

\noindent As described in Section 2.2.1, the density-based clustering algorithm is very promising to automate the analysis process. Its first step is to find good cluster centers in the decision graph which is composed of local density $\rho$ and $\delta$ the shortest distance to point with higher density. Seeing that the $\rho$ and $\delta$ of are all sensitive to the choice of criterion distance $d_c$, it is reasonable to vary $d_c$ when plotting decision graph. A DC value which defines the neighbor rate between 1 and 2 percent are used as the minimal unit to vary the criterion distance.\\
\begin{figure}[htbp]
\captionsetup{justification=centering}
\centering
\includegraphics[width=0.7\textwidth]{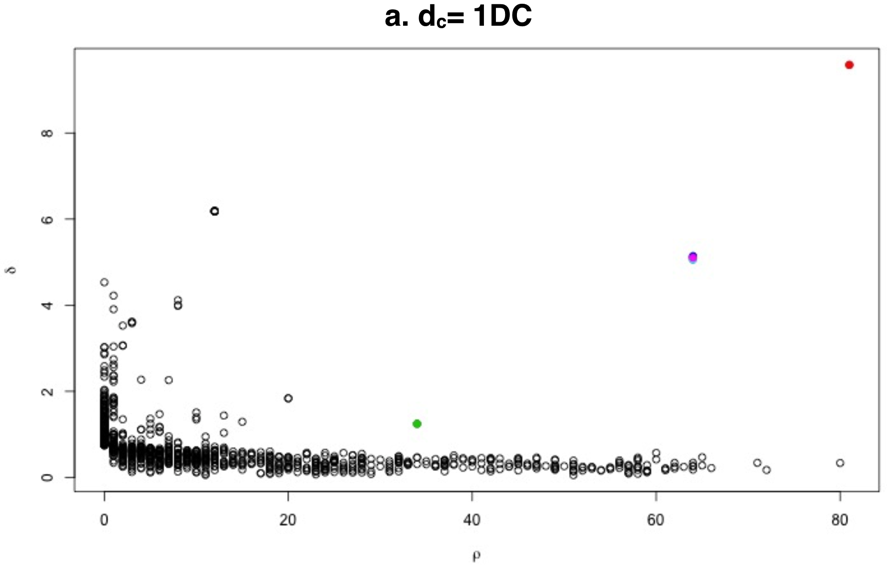}
\includegraphics[width=0.7\textwidth]{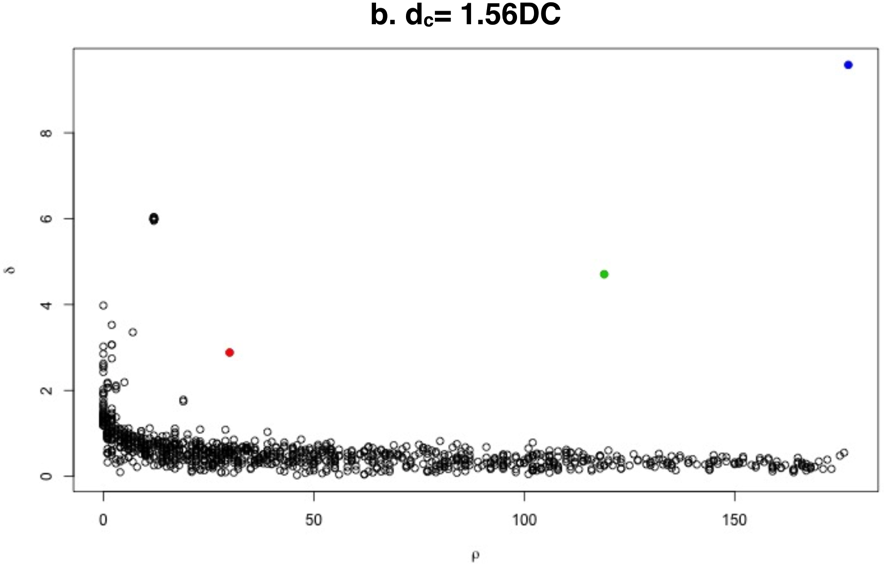}
\includegraphics[width=0.7\textwidth]{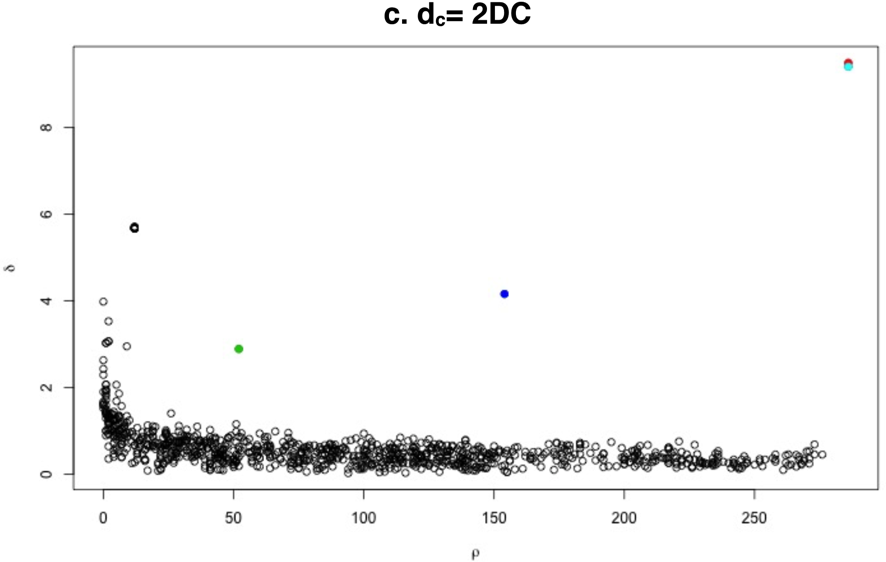}
\caption{\label{fig:decisiongraph}Decision graph of $\sum$5(210) when the criterion distance is equal to (a)1DC, (b)1.56DC, (c)2DC.}
\end{figure}

\begin{figure}[htbp]
%\captionsetup{justification=centering}
\centering
\includegraphics[width=1.0\textwidth]{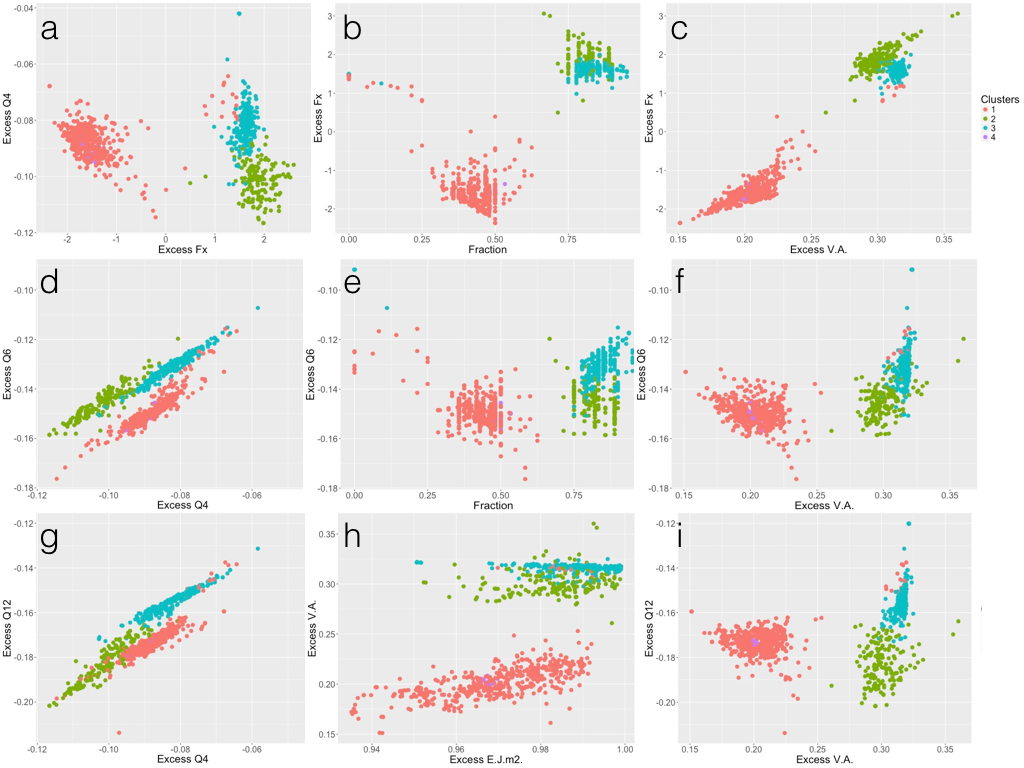}
\caption{\label{fig:good210}The clustering of GB structures in $\sum$5(210) system on different feature maps (a) Excess $Q_4$ vs Excess Stress $F_X$, (b)  Excess Stress $F_X$ vs Atomic Fraction, (c) Excess Stress $F_X$ vs Excess volume, (d) Excess $Q_6$ vs Excess $Q_4$, (e) Excess $Q_6$ vs Atomic Fraction, (f) Excess $Q_6$ vs Excess Volume, (g) Excess $Q_{12}$ vs Excess $Q_4$, (h) Excess Volume vs Excess Energy, (i) Excess $Q_{12}$ vs Excess Volume.}
\end{figure}

\noindent The Fig~\ref{fig:decisiongraph} shows the decision graphs of $\sum$5(210) when the criterion distance is set to be 1DC, 1.56DC and 2DC respectively. Only when 2DC is used to compute the $\rho$ and $\delta$, we find reasonable cluster centroids with higher local density and longer distance to points of more neighbors. According to these decision graphs, 2DC is selected to evaluate the $\delta$ and $\rho$ of every data point. The group is thereafter assigned for each point.\\
\\
The clustering behavior of $\sum$5(210) system can be further explored by visualizing these groups of points on different 2D feature maps. By looking the clustering performance of various feature pairs, we could decide good order parameters to distinguish different GB structures.\\
\\
We defined the GB structures in a 8-dimensional feature space. There are 28 different ways to map these structures into 2-dimensional subspaces. For most of feature pairs, we see very impressive clustering results for $\sum$5(210) system. Some of them are listed in Fig~\ref{fig:good210}. Comparing Fig~\ref{fig:good210}(c) and Fig~\ref{fig:210FxV}, it is clear to see that the density based algorithm yields very successful clustering results, which is as good as eye detection. In $\sum$29(520) system, we also see these encouraging results. What these results imply is that these feature are promising order parameters to cluster system with clustering behavior. This could greatly reduce the work load to analyze and define every structure manually. Most of the high angle GB structures show this intense clustering behaviors while for many lower tilt-angle system, very few feature pairs are effective to distinguish different GB structures. The latter point will be discussed in detail in the following section.

\subsection{Clustering of Grain Boundaries with Low Tilt Angle}
\begin{figure}[htbp]
%\captionsetup{justification=centering}
\centering
\includegraphics[width=1.0\textwidth]{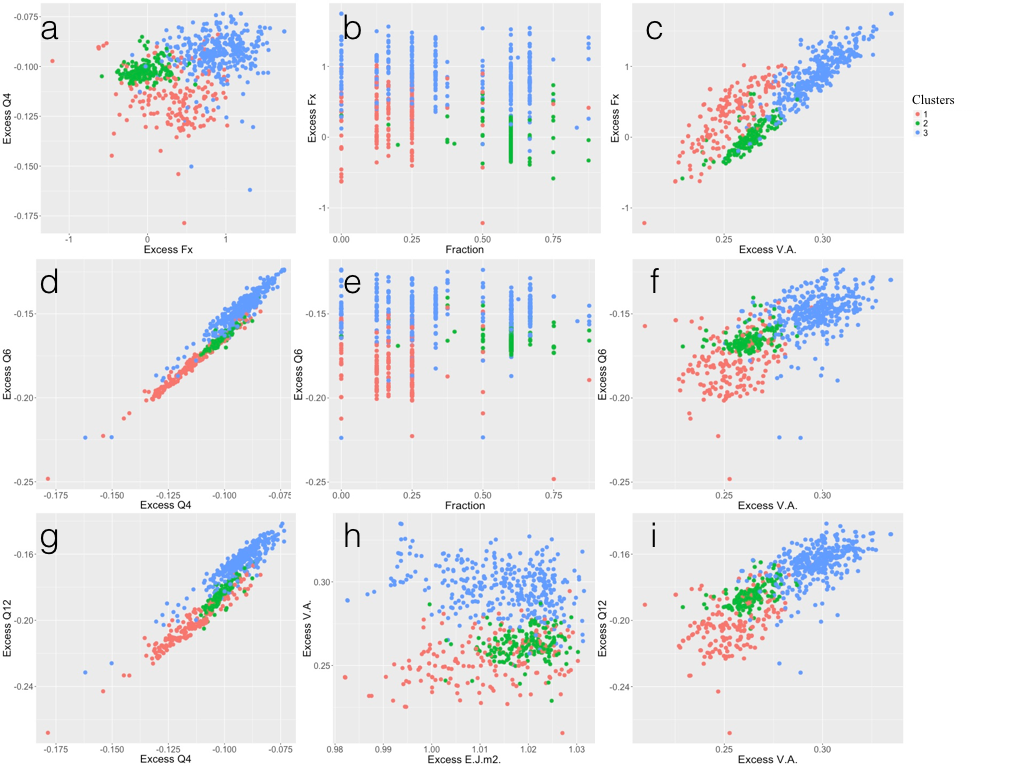}
\caption{\label{fig:good520}The clustering of GB structures in $\sum$29(520) system on different feature maps (a) Excess $Q_4$ vs Excess Stress $F_X$, (b)  Excess Stress $F_X$ vs Atomic Fraction, (c) Excess Stress $F_X$ vs Excess volume, (d) Excess $Q_6$ vs Excess $Q_4$, (e) Excess $Q_6$ vs Atomic Fraction, (f) Excess $Q_6$ vs Excess Volume, (g) Excess $Q_{12}$ vs Excess $Q_4$, (h) Excess Volume vs Excess Energy, (i) Excess $Q_{12}$ vs Excess Volume.}
\end{figure}

\noindent The $\sum$29(520) system has close tilt angle to $\sum$5(210) system. The tilt angles are 53.13$^\circ$ and 43.60$^\circ$ respectively for $\sum$29(520) and $\sum$5(210) system. We also observe satisfactory but not perfect clustering results in $\sum$29(520) system as shown in Fig~\ref{fig:good520}. However, the performance of excess energy and atomic fraction pair is not as ideal as that in $\sum$5(210) system.\\

\noindent As mentioned in previous section, most feature pairs are no longer good order parameters for lower angle system like $\sum$5(310)($\theta$=36.87$^\circ$), $\sum$17(410)($\theta$=28.07$^\circ$) and {\it etc.}. Usually the clustering are not satisfactory for the system with tilt angle under 36.87$^\circ$. The failure of these feature pairs comes from the disordered structures or transitory structures existing in these low tilt angle system.

\begin{figure}[htbp]
%\captionsetup{justification=centering}
\centering
\includegraphics[width=1.0\textwidth]{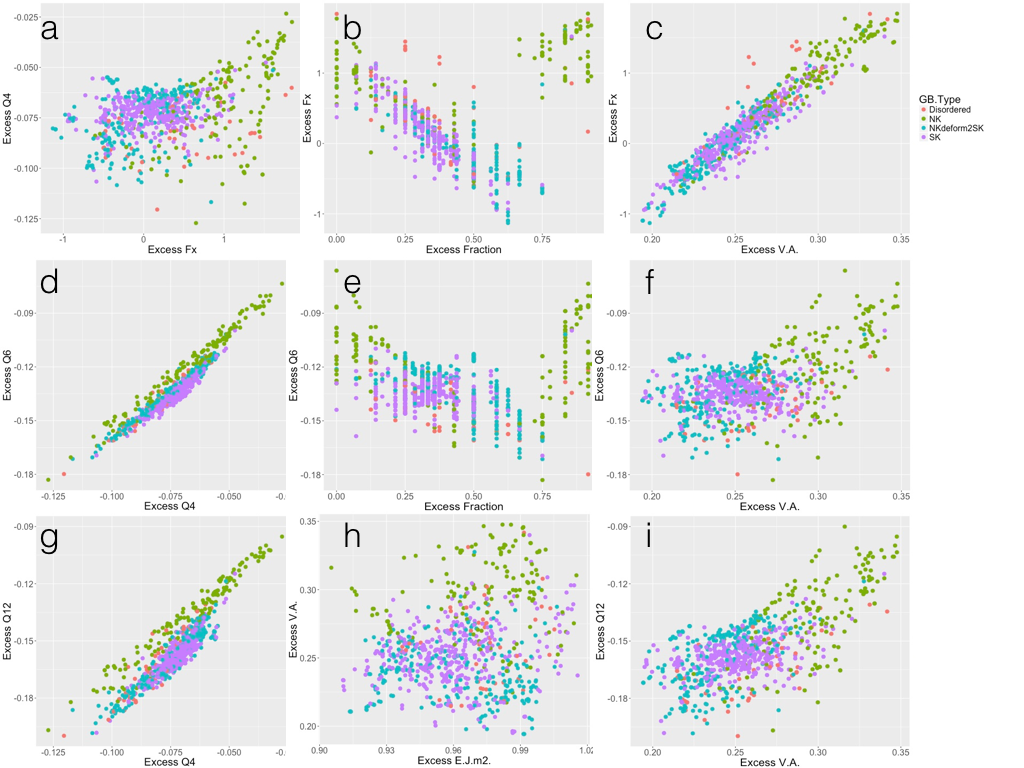}
\caption{\label{fig:good310}The clustering of GB structures in $\sum$5(310) system on different feature maps (a) Excess $Q_4$ vs Excess Stress $F_X$, (b) Excess Stress $F_X$ vs Atomic Fraction, (c) Excess Stress $F_X$ vs Excess volume, (d) Excess $Q_6$ vs Excess $Q_4$, (e) Excess $Q_6$ vs Atomic Fraction, (f) Excess $Q_6$ vs Excess Volume, (g) Excess $Q_{12}$ vs Excess $Q_4$, (h) Excess Volume vs Excess Energy, (i) Excess $Q_{12}$ vs Excess Volume.}
\end{figure}

\begin{figure}[htbp!]
\captionsetup{justification=centering}
\centering
\includegraphics[width=0.4\textwidth]{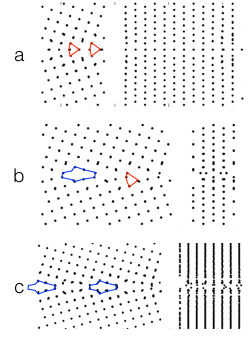}
\caption{\label{fig:310struc}The GB structures in $\sum$5(310) system (a) Kite Family (NK or K), (b) NKdeform2SK , (c) Split Kite family (SK).}{}
\end{figure}

\noindent In Fig~\ref{fig:good310}, we mapped the structures determined by eyes detection on 2-dimensional feature space. The clustering behavior is much less clear than that in $\sum$5(210) system. Only $Q_6$ and $Q_4$ well classified all of three GB structures, while the structure labeled as `NKdeform2SK' overlaps with Split Kite (SK) structure. The `NKdeform2SK' structure is actually a transitory structure between Kite family (NK) and Split Kite family (SK) according to Fig~\ref{fig:310struc} since its structure unit can be regarded as either deformed SK unit or deformed NK unit. According to overlapping between SK and NKdeform2SK structure in Fig~\ref{fig:good310}, the NKdeform2SK structure is closer to SK structure in the feature space, which implies their similarities in structural and energetic properties.\\

\begin{figure}[htbp!]
\captionsetup{justification=centering}
\centering
\includegraphics[width=0.8\textwidth]{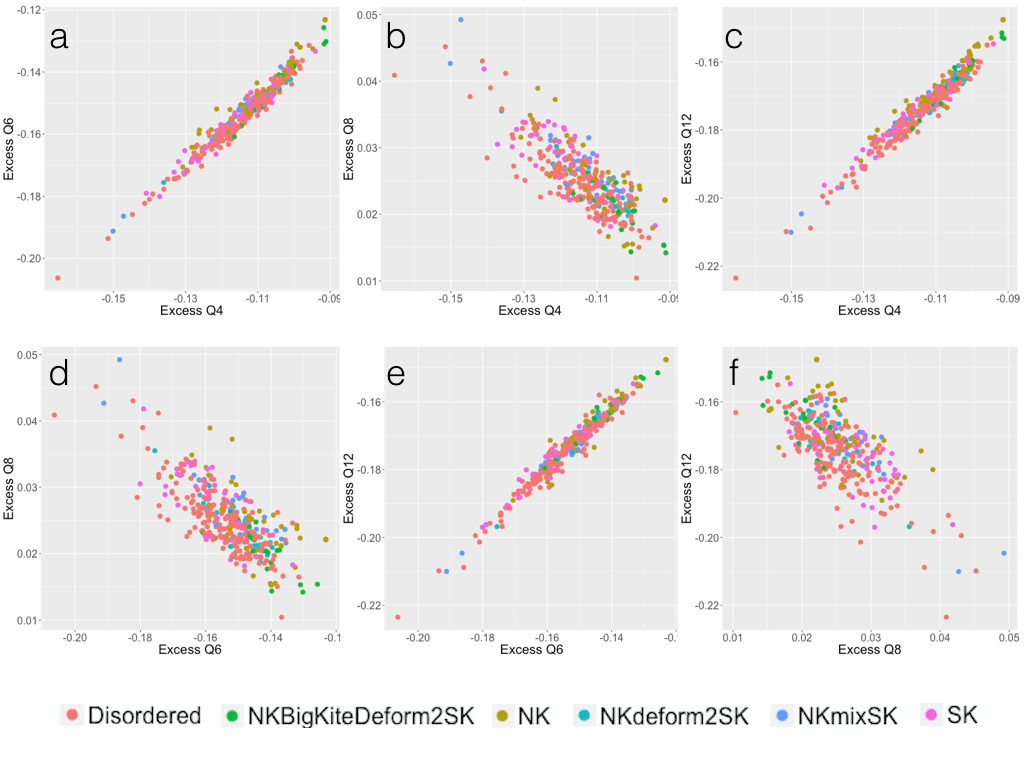}
\caption{\label{fig:good610}The GB structures in $\sum$5(310) system (a) Kite Family (NK or K), (b) NKdeform2SK , (c) Split Kite family (SK).}
\end{figure}

\begin{figure}[htbp!]
\captionsetup{justification=centering}
\centering
\includegraphics[width=0.8\textwidth]{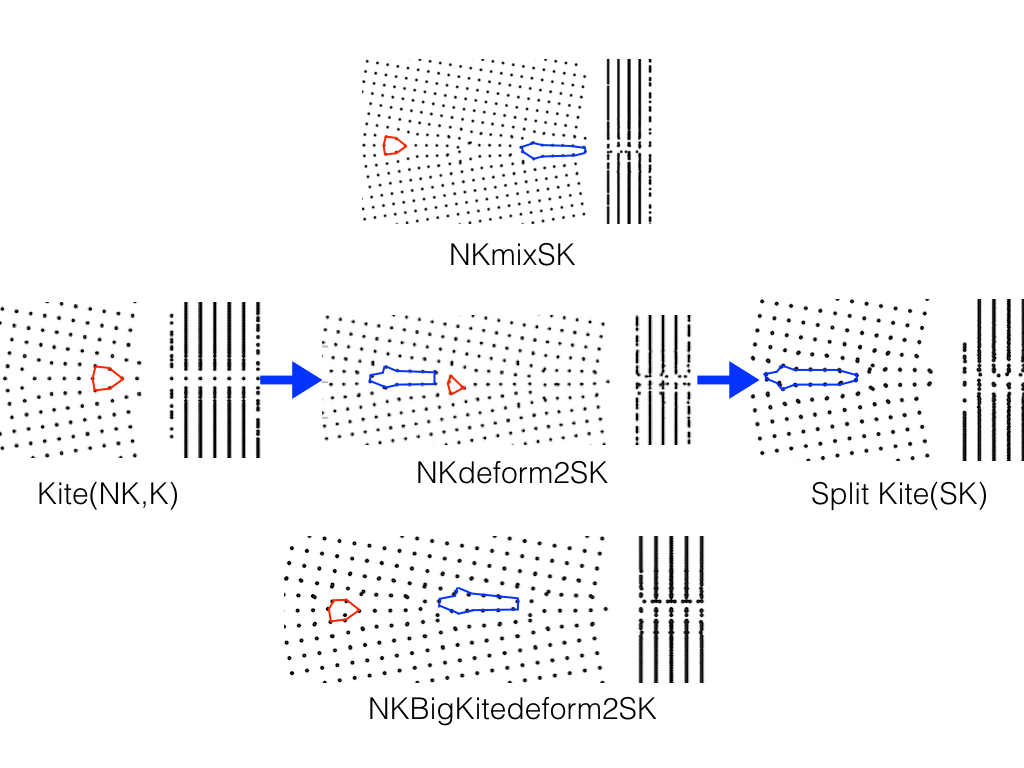}
\caption{\label{fig:610struc}The GB structures in $\sum$37(610) system Kite Family (NK or K) in the left , NKmixSK, NKdeform2SK and NKBigKitedeform2SK in the middle while Split Kite family (SK) in the right.}
\end{figure}

\noindent For system with lower tilt angle like $\sum$37(610) ($\theta$=18.92$^\circ$) and $\sum$25(710) ($\theta$=16.26$^\circ$),  the structures are even more complex. The complexity rises from the existence of mixed Kite and Split Kite structures and the transitory structures between these two kite families. These structures are usually disordered or partially ordered compare two kite families. In feature space, they usually overlaps with the two kite families for their partial structural similarities. These overlapping leads to the failure of clustering in current feature space. Therefore we hardly see clear clustering results when mapping them onto planes composed by any feature pairs. The Steinhardt Order Parameters, which is effective for $\sum$5(310), fail in the system of $\sum$37(610) as shown in Fig~\ref{fig:610struc} since the disordered and transitory structures like `NKdeform2SK', `NKmixSK', `NKBigKitedeform2SK' in Fig overlaps with NK and SK and blur the boundary between the two kite families.\\

\section{Discussions}
The development of this data pipeline helps understand the grain boundary structures and properties. Atomic arrangements in Cartesian coordinate system pose challenges to directly categorize a large number of structures. The feature engineering part of this data pipeline performs dimension reduction for each structures by mapping each structure into a space of 8 dimensions. Furthermore, the clustering analysis part of this data pipeline group these structures according to their distribution in the new space. A density based clustering algorithm\cite{rodriguez2014clustering} is used here to efficiently to find the group centroids and assign each structures.\\

\noindent The results are very intriguing and insightful. Structures are clustered into different groups very well and structures in different groups show different structure units at grain boundary region. These structures units are generalized as Kite, Split Kite, Filled and Extended Kite families respectively. However, when the tilt angle gets smaller, which means the grain boundary region is more compressed, more and more mixed and disordered kite families dominate the grain boundary structures. Some of them are the mixture of Normal Kite and Split Kite or intermediates of these two families. Generally, grain boundary structures have more diversity in lower tilt angle system while structures become more ordered in higher tilt angle system.\\

\noindent The current clustering is performed over structures at ambient pressure and 0~K and yields several different grain boundary families. It will be very interested to investigate grain boundary structures under different temperatures, which will reveal the phase diagram information of Grain Boundaries. It will also be a very fascinating topic to research on the evolution between different grain boundary structures with newly developed method like Transition Path Sampling method~\cite{bolhuis2002transition}. My recent work imply the potential to adapt this method to investigate the bulk phase transition~\cite{doi:10.1063/1.4983167}.\\
\\
\noindent The parallel K-means algorithm function is implemented here and left for future big data application. For this grain boundary problem, the density based clustering algorithm already yields very satisfying results efficiently.

\section{Acknowledgments}
The work was funded by the Laboratory Directed Research and Development Program at Lawrence Livermore National Laboratory(LLNL) under project tracking code 17-LW-012. Prof. Qiang Zhu at University of Las Vegas, Nevada predicted grain boundary structures with evolutionary algorithm. Dr. Timofey Frolov at LLNL provided helpful suggestions on the feature properties calculations to grain boundary structures. Prof. Norman S. Matloff at University of California, Davis advised me during this project.

\section{Running Demo}
The running of the data pipeline is given to Cu $\sum$29(520) as a demo.\\
\\
Enter the computation directory.
\begin{lstlisting}
******************* Step 1. Run the Properties Calculation *******************
************************    1.1 Run the Calculation   ************************    
>> cd Cu520
\end{lstlisting}

\noindent Rescale the dimension and atoms coordinates of predicted structures to be consistent with correct lattice parameter 3.615 for Cu face center cubic cell of 4 atoms with a=b=c and $\alpha$=$\beta$=$\gamma$=90$^\circ$. The \textbf{LAMMPS\_backup} is the input structure directory and the \textbf{LAMMPS\_backup\_new} is the rescaled structure output directory.
\begin{lstlisting}
>> nohup ./convert.py LAMMPS_backup LAMMPS_backup_new 3.615 5 2 0 & 
\end{lstlisting}

\noindent Perform the properties calculation over the rescaled structures, you can track the calculation progress from \textbf{progress} file in current directory while all the structures's properties are collected in \textbf{STRUCTURES.dat} file. Here the potential file \textbf{Cu01.eam.alloy.txt} is assigned for atomic interactions used in LAMMPS structure local optimization.\\
\textbf{NOTE:} During this calculation, you can check the `progress' file to track the progress.
\begin{lstlisting}
>> nohup ./pipe.py LAMMPS_backup_new Cu01.eam.alloy.txt 5 2 0 fcc Cu &
\end{lstlisting}

\begin{lstlisting}
**********    1.2 SUMMARY of Results    **********   
**** 1.2.1 about the calc folder  ****
\end{lstlisting}
After finishing the above calculation, there will be 10 calculation folder, \textbf{calc1}, \textbf{cal2}, \textbf{calc3}, $\ldots$, \textbf{calc10}. For structure LAMMPS\_\#X, it will be computed under calc\#T, where T = X $\%$10 if X$\%$10 $\neq$ 0 and T = 10 if X$\%$10 == 0. For example, LAMMMPS\_0001 is calcualted in calc1 folder and LAMMPS\_0010 is calculated in \textbf{calc10} folder. 
\noindent In each calculation folder, you will find the following calculation files. Let us look into the calculation to structure indexed by 579 under \textbf{calc9}.
\begin{lstlisting}
>> ls -al ./calc9/*0579*
-rw-r--r--@ 1 BingxiLi  staff     759 Dec  5 17:19 ./calc9/SUBMIT_lammps_0579
-rw-r--r--@ 1 BingxiLi  staff      84 Dec  5 17:19 ./calc9/lammps_0579.error
-rw-r--r--@ 1 BingxiLi  staff    6005 Dec  5 17:19 ./calc9/lammps_0579.in
-rw-r--r--@ 1 BingxiLi  staff    7810 Dec  5 17:19 ./calc9/lammps_0579.output
-rw-r--r--@ 1 BingxiLi  staff  324351 Dec  5 17:19 ./calc9/lammps_0579.struc
-rw-r--r--@ 1 BingxiLi  staff   36772 Dec  5 17:19 ./calc9/lammps_0579_low.struc
-rw-r--r--@ 1 BingxiLi  staff   36624 Dec  5 17:19 ./calc9/lammps_0579_up.struc
\end{lstlisting}

\noindent Here is the explanation:
\begin{enumerate}
	\item SUBMIT\_lammps\_0579: show how to SUBMIT the calculation. If you want to repeat the calculatio, please run 'sbatch ./calc9/SUBMIT\_lammps\_0579'.
	\item lammps\_0579.in: the lammps input for the properties calculation to structure 0579.
	\item lammps\_0579.error and lammps\_0579.output: lammps output files.
	\item *.Struc: lammps dump file generated at the end of lammps calculation. It contains x, y, z coordinates and different properties for each atoms in the gb structure. More in details:
	\begin{enumerate}
		\item lammps\_0579.struc: the coordinates and properties information for all the atoms in the structures
		\item lammps\_0579\_low.struc: the coordinates and properties information for atoms in the selected bulk region in lower layer.
		\item lammps\_0579\_up.struc: the coordinates and properties information for atoms in the selected bulk region in upper layer.
	\end{enumerate}
\end{enumerate}
\noindent \textbf{NOTE:} The bulk region in lower and upper layer here use the region that is 10A far from their surface. This choice are validated effective in Step 2.\\
\noindent The \textbf{STRUCTURES.dat} listed the excess properties for each structures. Here is an example of its content. 
\begin{lstlisting}
**** 1.2.2 about the STRUCTURES.dat ****
>> head -2 STRUCTURES.dat
   Structures    Fraction  E[J/m2]     V[A]        Fx          Fy       Fz        Q4         Q6         Q8         Q12
   lammps_0001   0.00000   0.98299   0.28863   1.37559   0.51736   -0.00640   -0.09975   -0.14718   0.01992   -0.16804
\end{lstlisting}

\noindent \\The following shows how to verify the properties results by comparing the neighborhood of grain boundary region to ideal bulk. When neighborhoods have close value to ideal bulk which means the structure is well relaxed. 
\begin{lstlisting}
***********************  Step 2. Validate the Results  *********************** 
************************    2.1 Run the Calculation   ************************
\end{lstlisting}
Generates \textbf{AVGSTDS\_lowbulk.dat} and \textbf{AVGSTDS\_upbulk.dat} to store two side neighborhood regions' property results. Plot over \textbf{AVGSTDS\_lowbulk.dat} and \textbf{AVGSTDS\_upbulk.dat} and figures will be saved to \textbf{AVGSTDS\_lowbulk\_plots} and \textbf{AVGSTDS\_upbulk\_plots} respectively.
\begin{lstlisting}  
>> ./avgstd_lowbulk.py LAMMPS_backup_new/			
>> ./avgstd_upbulk.py LAMMPS_backup_new/
>> ./avgstd_bulkplots.py AVGSTDS_lowbulk.dat AVGSTDS_lowbulk_plots AVGSTDS_upbulk.dat AVGSTDS_upbulk_plots
\end{lstlisting}
\noindent Sometimes there will be great discrepancy between neighborhood bulk and ideal bulk, which means the grain boundary is either poorly relaxed or falsely assigned. Usually the latter reason is more likely since the previous calculation assign the grain boundary region artificially. Namely, we simply set it as a fixed region within a certain distance range in previous run. In the case where discrepancies arise, we need to rerun the properties with following commands. By doing this, we could dynamically assign accurate grain boundary according to energy and Steinhardt Order Parameters. The \textbf{rerun.py} generates \textbf{STRUCTURESnew.dat} to store properties of newly assigned grain boundary regions and \textbf{AVGSTDS\_lowbulknew.dat}, \textbf{AVGSTDS\_upbulknew.dat} to store properties for new neighborhoods. By plotting and comparing, we could see the robustness of this way to finding accurate grain boundary regions.
\begin{lstlisting}
>> nohup ./rerun.py ./LAMMPS_backup_new/ 5 2 0 &	
>> ./avgstd_bulkplots.py AVGSTDS_lowbulknew.dat AVGSTDS_lowbulk_plots_new AVGSTDS_upbulknew.dat AVGSTDS_upbulk_plots_new
\end{lstlisting}

\noindent \\The following shows simple clustering analysis operations.
\begin{lstlisting}$
***********************   Step 3. Analyze the Results  ***********************
$ Rscript DClust.R ../STRUCTURESnew.dat 1.56 20 2
\end{lstlisting}
The script \textbf{DClust.R} takes 4 arguments. The first one \textbf{STRUCTURESnew.dat} is the properties file and the second one sets how many times of default distance criterion, which is selected to make neighbor rates between 1\%--2\%. The third and fourth one assign the $\rho$ and $\delta$ criterion to select centroids on decision graph. Clustering results visualized on different 2--D plots will be output to \textbf{DClustAnalysis\_plots/} directory and the corresponding decision graph can also be found as \textbf{Decision Graph} in that directory.
% \begin{thebibliography}{99}

% \bibitem{melissinos}
% A.~C. Melissinos and J. Napolitano, \textit{Experiments in Modern Physics},
% (Academic Press, New York, 2003).

% \bibitem{Cyr}
% N.\ Cyr, M.\ T$\hat{e}$tu, and M.\ Breton,
% % "All-optical microwave frequency standard: a proposal,"
% IEEE Trans.\ Instrum.\ Meas.\ \textbf{42}, 640 (1993).

% \bibitem{Wiki} \emph{Expected value},  available at
% \texttt{http://en.wikipedia.org/wiki/Expected\_value}.

% \end{thebibliography}

\newpage
\singlespacing
\bibliographystyle{plain}
\bibliography{./chapter1.bib}

\begin{thebibliography}{10}

\bibitem{dillon2007complexion}
Shen~J Dillon, Ming Tang, W~Craig Carter, and Martin~P Harmer.
\newblock Complexion: A new concept for kinetic engineering in materials
  science.
\newblock {\em Acta Materialia}, 55(18):6208--6218, 2007.

\bibitem{CANTWELL20141}
Patrick~R. Cantwell, Ming Tang, Shen~J. Dillon, Jian Luo, Gregory~S. Rohrer,
  and Martin~P. Harmer.
\newblock Grain boundary complexions.
\newblock {\em Acta Materialia}, 62:1 -- 48, 2014.

\bibitem{RICKMAN201388}
J.M. Rickman, H.M. Chan, M.P. Harmer, and J.~Luo.
\newblock Grain-boundary layering transitions in a model bicrystal.
\newblock {\em Surface Science}, 618:88 -- 93, 2013.

\bibitem{tang2006grain}
Ming Tang, W~Craig Carter, and Rowland~M Cannon.
\newblock Grain boundary transitions in binary alloys.
\newblock {\em Physical review letters}, 97(7):075502, 2006.

\bibitem{tang2006diffuse}
Ming Tang, W~Craig Carter, and Rowland~M Cannon.
\newblock Diffuse interface model for structural transitions of grain
  boundaries.
\newblock {\em Physical Review B}, 73(2):024102, 2006.

\bibitem{luo1999origin}
Jian Luo, Haifeng Wang, and Yet-Ming Chiang.
\newblock Origin of solid-state activated sintering in bi2o3-doped zno.
\newblock {\em Journal of the American Ceramic Society}, 82(4):916--920, 1999.

\bibitem{luo2011role}
Jian Luo, Huikai Cheng, Kaveh~Meshinchi Asl, Christopher~J Kiely, and Martin~P
  Harmer.
\newblock The role of a bilayer interfacial phase on liquid metal
  embrittlement.
\newblock {\em Science}, 333(6050):1730--1733, 2011.

\bibitem{kaplan2015mechanism}
Wayne~D Kaplan.
\newblock The mechanism of crystal deformation.
\newblock {\em Science}, 349(6252):1059--1060, 2015.

\bibitem{kuzmina2015linear}
Margarita Kuzmina, Michael Herbig, Dirk Ponge, Stefanie Sandl{\"o}bes, and
  Dierk Raabe.
\newblock Linear complexions: Confined chemical and structural states at
  dislocations.
\newblock {\em Science}, 349(6252):1080--1083, 2015.

\bibitem{baram2011nanometer}
Mor Baram, Dominique Chatain, and Wayne~D Kaplan.
\newblock Nanometer-thick equilibrium films: the interface between
  thermodynamics and atomistics.
\newblock {\em Science}, 332(6026):206--209, 2011.

\bibitem{rheinheimer2015non}
Wolfgang Rheinheimer and Michael~J Hoffmann.
\newblock Non-arrhenius behavior of grain growth in strontium titanate: new
  evidence for a structural transition of grain boundaries.
\newblock {\em Scripta Materialia}, 101:68--71, 2015.

\bibitem{dillon2016importance}
Shen~J Dillon, Kaiping Tai, and Song Chen.
\newblock The importance of grain boundary complexions in affecting physical
  properties of polycrystals.
\newblock {\em Current Opinion in Solid State and Materials Science},
  20(5):324--335, 2016.

\bibitem{rohrer2016role}
Gregory~S Rohrer.
\newblock The role of grain boundary energy in grain boundary complexion
  transitions.
\newblock {\em Current Opinion in Solid State and Materials Science},
  20(5):231--239, 2016.

\bibitem{merkle1987atomic}
KL~Merkle and David~J Smith.
\newblock Atomic structure of symmetric tilt grain boundaries in nio.
\newblock {\em Physical Review Letters}, 59(25):2887, 1987.

\bibitem{park2003singular}
Chan~Woo Park, Duk~Yong Yoon, John~E Blendell, and Carol~A Handwerker.
\newblock Singular grain boundaries in alumina and their roughening transition.
\newblock {\em Journal of the American Ceramic Society}, 86(4):603--11, 2003.

\bibitem{frolov2015segregation}
Timofey Frolov, Mark Asta, and Yuri Mishin.
\newblock Segregation-induced phase transformations in grain boundaries.
\newblock {\em Physical Review B}, 92(2):020103, 2015.

\bibitem{frolov2016phase}
T~Frolov, M~Asta, and Y~Mishin.
\newblock Phase transformations at interfaces: Observations from atomistic
  modeling.
\newblock {\em Current Opinion in Solid State and Materials Science},
  20(5):308--315, 2016.

\bibitem{oganov2006crystal}
Artem~R Oganov and Colin~W Glass.
\newblock Crystal structure prediction using ab initio evolutionary techniques:
  Principles and applications.
\newblock {\em The Journal of chemical physics}, 124(24):244704, 2006.

\bibitem{zhou2014semimetallic}
Xiang-Feng Zhou, Xiao Dong, Artem~R Oganov, Qiang Zhu, Yongjun Tian, and
  Hui-Tian Wang.
\newblock Semimetallic two-dimensional boron allotrope with massless dirac
  fermions.
\newblock {\em Physical Review Letters}, 112(8):085502, 2014.

\bibitem{zhu2013evolutionary}
Qiang Zhu, Li~Li, Artem~R Oganov, and Philip~B Allen.
\newblock Evolutionary method for predicting surface reconstructions with
  variable stoichiometry.
\newblock {\em Physical Review B}, 87(19):195317, 2013.

\bibitem{zhu2014predicting}
Qiang Zhu, Vinit Sharma, Artem~R Oganov, and Ramamurthy Ramprasad.
\newblock Predicting polymeric crystal structures by evolutionary algorithms.
\newblock {\em The Journal of chemical physics}, 141(15):154102, 2014.

\bibitem{lyakhov2013new}
Andriy~O Lyakhov, Artem~R Oganov, Harold~T Stokes, and Qiang Zhu.
\newblock New developments in evolutionary structure prediction algorithm
  uspex.
\newblock {\em Computer Physics Communications}, 184(4):1172--1182, 2013.

\bibitem{von2006structures}
S~Von~Alfthan, PD~Haynes, K~Kaski, and AP~Sutton.
\newblock Are the structures of twist grain boundaries in silicon ordered at 0
  k?
\newblock {\em Physical review letters}, 96(5):055505, 2006.

\bibitem{chua2010genetic}
Alvin~Ls Chua, Nicole~A Benedek, Lin Chen, Mike~W Finnis, and Adrian~P Sutton.
\newblock A genetic algorithm for predicting the structures of interfaces in
  multicomponent systems.
\newblock {\em Nature materials}, 9(5):418, 2010.

\bibitem{zhang2009finding}
Jian Zhang, Cai-Zhuang Wang, and Kai-Ming Ho.
\newblock Finding the low-energy structures of si [001] symmetric tilted grain
  boundaries with a genetic algorithm.
\newblock {\em Physical Review B}, 80(17):174102, 2009.

\bibitem{steinhardt1983bond}
Paul~J Steinhardt, David~R Nelson, and Marco Ronchetti.
\newblock Bond-orientational order in liquids and glasses.
\newblock {\em Physical Review B}, 28(2):784, 1983.

\bibitem{frolov2012thermodynamics}
T~Frolov and Y~Mishin.
\newblock Thermodynamics of coherent interfaces under mechanical stresses. i.
  theory.
\newblock {\em Physical Review B}, 85(22):224106, 2012.

\bibitem{plimpton1995fast}
Steve Plimpton.
\newblock Fast parallel algorithms for short-range molecular dynamics.
\newblock {\em Journal of computational physics}, 117(1):1--19, 1995.

\bibitem{rodriguez2014clustering}
Alex Rodriguez and Alessandro Laio.
\newblock Clustering by fast search and find of density peaks.
\newblock {\em Science}, 344(6191):1492--1496, 2014.

\bibitem{le1972proceedings}
Lucien Le~Cam, Jerzy Neyman, and Elizabeth~L Scott.
\newblock {\em Proceedings of the Sixth Berkeley Symposium on Mathematical
  Statistics and Probability: Held at the Statistical Laboratory, University of
  California, June 21-July 18, 1970}, volume~2.
\newblock Univ of California Press, 1972.

\bibitem{mclachlan1997wiley}
Geoffrey~J McLachlan and Thriyambakam Krishnan.
\newblock Wiley series in probability and statistics.
\newblock {\em The EM Algorithm and Extensions, Second Edition}, pages
  361--369, 1997.

\bibitem{inaba1994applications}
Mary Inaba, Naoki Katoh, and Hiroshi Imai.
\newblock Applications of weighted voronoi diagrams and randomization to
  variance-based k-clustering.
\newblock In {\em Proceedings of the tenth annual symposium on Computational
  geometry}, pages 332--339. ACM, 1994.

\bibitem{christopher2008introduction}
D~Manning Christopher, Raghavan Prabhakar, and SCH{\"U}TZE Hinrich.
\newblock Introduction to information retrieval.
\newblock {\em An Introduction To Information Retrieval}, 151:177, 2008.

\bibitem{arthur2006slow}
David Arthur and Sergei Vassilvitskii.
\newblock How slow is the k-means method?
\newblock In {\em Proceedings of the twenty-second annual symposium on
  Computational geometry}, pages 144--153. ACM, 2006.

\bibitem{submittedbl}
Qiang Zhu, Amit Samanta, Bingxi Li, Robert~E. Rudd, and Timofey Frolov.
\newblock Predicting phase behavior of grain boundaries with evolutionary
  search and machine learning.
\newblock {\em submitted}, 2017.

\bibitem{bolhuis2002transition}
Peter~G Bolhuis, David Chandler, Christoph Dellago, and Phillip~L Geissler.
\newblock {\em Ann. ReV. Phys. Chem.}, 53(1):291--318, 2002.

\bibitem{doi:10.1063/1.4983167}
Bingxi Li, Guangrui Qian, Artem~R. Oganov, Salah~Eddine Boulfelfel, and Roland
  Faller.
\newblock Mechanism of the fcc-to-hcp phase transformation in solid ar.
\newblock {\em The Journal of Chemical Physics}, 146(21):214502, 2017.

\end{thebibliography}

\end{document}